\documentclass[aps,pre,groupedaddress,showpacs,preprint]{revtex4}
\usepackage{graphicx}
\usepackage[dvips]{epsfig}
\usepackage{dcolumn}
\usepackage{subfigure}%
\usepackage{bm}
\usepackage{amssymb}
\usepackage{amsmath}

\begin{document}
\title{Stochastic growth equations on growing domains}

\author{Carlos Escudero}

\affiliation{Instituto de Ciencias Matem\'{a}ticas, Consejo Superior
de Investigaciones Cient\'{\i}ficas, C/ Serrano 123, 28006 Madrid,
Spain}

\begin{abstract}
The dynamics of linear stochastic growth equations on growing
substrates is studied. The substrate is assumed to grow in time
following the power law $t^\gamma$, where the growth index $\gamma$
is an arbitrary positive number. Two different regimes are clearly
identified: for small $\gamma$ the interface becomes correlated, and
the dynamics is dominated by diffusion; for large $\gamma$ the
interface stays uncorrelated, and the dynamics is dominated by
dilution. In this second regime, for short time intervals and
spatial scales the critical exponents corresponding to the
non-growing substrate situation are recovered. For long time
differences or large spatial scales the situation is different.
Large spatial scales show the uncorrelated character of the growing
interface. Long time intervals are studied by means of the
auto-correlation and persistence exponents. It becomes apparent that
dilution is the mechanism by which correlations are propagated in
this second case.
\end{abstract}


\maketitle

\section{Introduction}

Fluctuating interfaces have been the object of study of many
different works over the last decades. Together with the possible
technological applications that their understanding may bring, as
for instance in thin film industry, there is a genuine theoretical
interest in unveiling their dynamical properties. This is so to the
extend that the Kardar-Parisi-Zhang (KPZ) equation \cite{kpz}, one
of the most influential models for surface growth, is being
currently considered as a prototypical model of nonequilibrium
dynamics.

Usually, stochastic equations modeling surface growth have been
studied in static domains. On the other hand, some types of growing
interfaces, as for instance radial ones, present a domain size that
grows over time. The direct study of radial interfaces is
complicated by nonlinear effects, including the possibility of
instabilities affecting the radial symmetry
\cite{kapral,batchelor,escudero}. This suggests studying first the
dynamics of linear stochastic growth equations on growing domains.
Once the effect of substrate growth on the interface dynamics is
understood, one could move to the more complicated case of radial
growth.

Of course, considering linear growth equations has a limited
applicability to real physical systems, as important
nonlinearities are being neglected. Still, the detailed analysis
of linear stochastic growth equations has revealed important
physical properties of rough surfaces \cite{nissila}.
Additionally, the study of linear equations has served as a basis
for the approach to the more complicated nonlinear ones
\cite{tang}. In this sense, we expect that the results presented
here might bring useful insights into the dynamics of nonlinear
equations on growing domains.

Stochastic growth equations for radial interfaces have been
previously considered in the literature
\cite{kapral,batchelor,escudero,marsili,singha}. In \cite{singha},
radial interfaces with a linearly in time growing domain are
studied, and several dynamical quantities are calculated and
compared to the classical values. Under the approximations made in
this work, the only genuine radial effect that is being considered
is domain growth, while the possible nonlinear effects are
disregarded. This allows a perfect comparison among the results
presented there and the ones that we will introduce here. We will
also examine how decorrelation might appear in the growing interface
and what is the resulting large scale structure \cite{escudero}, how
the classical values of the critical exponents are recovered
\cite{krug2}, and in what precise limits this occurs
\cite{escudero2}.

The goal of this work is to put previous studies considering
linearly in time growing domains in a broader context. This will be
achieved by proposing an arbitrary power law growth model for the
domain. We will focus on rough interfaces, i. e., models for which
the growth exponent $\beta$ is strictly positive. The paper is
organized as follows: in section \ref{domains} we describe the
phenomenology of domain growth. In sec. \ref{ewdynamics} we focus on
the nonequilibrium dynamics of the Edwards-Wilkinson (EW) equation,
and in sec. \ref{general} we extend these results to the general
linear Langevin dynamics. In sec. \ref{temporal} the temporal
correlations of the fluctuating interface are studied, and in sec.
\ref{persitence} its persistence properties. In sec.
\ref{connection} the connection to radial growth is investigated,
and the conclusions of this work are drawn in sec.
\ref{conclusions}.

\section{Growing domains}
\label{domains}

In order to study the dynamics of stochastic growth equations on
growing domains we begin considering the EW equation \cite{EW},
which reads
\begin{equation}
\partial_t h=D\nabla^2 h + F + \xi({\bf y},t),
\end{equation}
where $\xi({\bf y},t)$ is a zero-mean Gaussian white noise which
correlation is
\begin{equation}
\langle \xi({\bf y},t)\xi({\bf y}',t') \rangle=\epsilon \delta({\bf
y}-{\bf y}')\delta(t-t'),
\end{equation}
$D$ is the diffusion constant, $F$ the constant deposition rate and
$\epsilon$ the noise intensity, all these parameters being positive.
To derive the EW equation on a growing domain we will follow the
theory introduced in \cite{crampin}, focused on reaction diffusion
dynamics on uniformly growing domains. We start considering the
conservation law in integral form
\begin{equation}
\frac{d}{dt}\int_{S_t} h({\bf y},t)d{\bf y}=\int_{S_t}\left[ -\nabla
\cdot {\bf j}+ \mathcal{F}({\bf y},t) \right]d{\bf y},
\end{equation}
where $S_t$ is the uniformly growing domain, ${\bf j}=-D\nabla h$ is
the current generated by diffusion, and $\mathcal{F}({\bf
y},t)=F+\xi({\bf y},t)$ is the EW growth mechanism. By applying
Reynolds transport theorem we find
\begin{equation}
\frac{d}{dt}\int_{S_t}h({\bf y},t)d{\bf
y}=\int_{S_t}\left[\partial_t h + \nabla \cdot ({\bf v}h) \right]
d{\bf y},
\end{equation}
where ${\bf v}({\bf y},t)$ denotes the flow velocity generated by
the growing domain. Valid as it is for any domain, the integral
conservation law may be expressed in the local form
\begin{equation}
\label{local}
\partial_t h+\nabla \cdot ({\bf v}h)=D\nabla^2h+\mathcal{F}({\bf y},t).
\end{equation}
In this equation we readily identify two new terms, the advection
one ${\bf v}\cdot \nabla h$, and the \emph{dilution} one $h \nabla
\cdot {\bf v}$. For every ${\bf y} \in S_t$, that has evolved from
${\bf y}_0 \in S_{t_0}$, we find ${\bf v}({\bf y},t)=\partial {\bf
y}/\partial t$. Let us now concentrate in one-dimensional
substrates and then move to higher dimensionalities. In this case
uniform growth translates into $y=g(t)y_0$, where $g(t)$ is a
temporal function such that $g(t_0)=1$. This yields
$v=y\dot{g}/g$, and thus
\begin{equation}
\partial_t h +\frac{\dot{g}}{g}\left( y\partial_y h + h \right)= D \partial_y^2 h + F + \xi(y,t).
\end{equation}
For a one-dimensional substrate $\left(0,L(t)\right)$, with
$L(t)=g(t)L_0$, we change the spatial coordinate $x = y L_0/L(t)$,
where $L_0=L(t_0)$, in order to map the problem into the interval
$\left( 0,L_0 \right)$. This transformation counterbalances
advection, and so the resulting equation reads
\begin{equation}
\frac{\partial h}{\partial t}=\left(\frac{L_0}{L(t)}\right)^2D
\frac{\partial^2 h}{\partial x^2}-\frac{\dot{g}}{g}h+F+ \sqrt{\frac{L_0}{L(t)}} \xi(x,t),
\end{equation}
where we have used the fact that the noise is delta correlated.
The dilution term has become $h \nabla \cdot {\bf v}=
-(\dot{g}/g)h$. It has been disregarded in reaction-diffusion
systems due to its irrelevance in this context \cite{crampin}, but
we will keep it here, where it will show its measurable effects on
the dynamics. Indeed, dilution has a transparent physical meaning:
as the substrate grows the deposited material becomes distributed
in a larger ($d-$dimensional) area. This matter redistribution
causes in turn the propagation of correlations, additionally to
diffusion, resulting in a different dynamical scenario as the
following sections will show. Now we assume that the growth
function adopts the power law form $g(t)=(t/t_0)^\gamma$, where
the growth index $\gamma \ge 0$, to find
\begin{equation}
\label{gamma} \frac{\partial h}{\partial
t}=\left(\frac{t_0}{t}\right)^{2\gamma} D \frac{\partial^2
h}{\partial x^2}-\frac{\gamma}{t}h+F+
\left(\frac{t_0}{t}\right)^{\gamma/2} \xi(x,t).
\end{equation}
The growth index $\gamma$ is a new degree of freedom of this
problem; it cannot be deduced from the other model parameters, and
has to be measured directly from the physical system under study.
Our next step is to assume no flux boundary conditions $\partial_x
h(0,t)=\partial_x h(L_0,t)=0$, both due to their physical
relevance and because they break translation invariance. It will
be interesting to see how translation invariance is recovered as a
consequence of decorrelation for rapidly growing domains. Lets
decompose the solution in the basis formed by the eigenfunctions
of the Laplacian on the domain under consideration
\begin{equation}
\label{basis} h(x,t)=\sum_{n=0}^{\infty} h_n(t) \cos \left(\frac{n
\pi x}{L_0}\right),
\end{equation}
to reduce Eq. (\ref{gamma}) to a stochastic differential equation for the different modes
\begin{equation}
\label{gammaf} \frac{d h_n}{dt}=-\frac{n^2 \pi^2}{L_0^2}
\left(\frac{t_0}{t}\right)^{2\gamma} D h_n-\frac{\gamma}{t}h_n+
\left(\frac{t_0}{t}\right)^{\gamma/2} \xi_n(t),
\end{equation}
if $n \neq 0$, and
\begin{equation}
\frac{d
h_0}{dt}=-\frac{\gamma}{t}h_0+F+\left(\frac{t_0}{t}\right)^{\gamma/2}\xi_0(t).
\end{equation}
In these equations $\xi_n(t)$ is a Gaussian random variable with zero mean and correlation given by
\begin{eqnarray}
\langle \xi_m(t)\xi_n(t') \rangle &=& \frac{2\epsilon}{L_0}\delta_{mn}\delta(t-t'), \qquad \mathrm{if} \qquad n,m \neq 0, \\
\langle \xi_0(t)\xi_n(t') \rangle &=& 0, \qquad \mathrm{if} \qquad n \neq 0, \\
\langle \xi_0(t)\xi_0(t') \rangle &=&
\frac{\epsilon}{L_0}\delta(t-t').
\end{eqnarray}

\section{Edwards-Wilkinson dynamics}
\label{ewdynamics}

We can straightforwardly derive the equation of motion for $\langle
h_0 \rangle$
\begin{equation}
\frac{d \langle h_0 \rangle}{dt}=-\frac{\gamma}{t} \langle h_0 \rangle + F,
\end{equation}
whose long time solution reads
\begin{equation}
\langle h_0 \rangle=\frac{F}{\gamma+1}t.
\end{equation}
For the second moment we find
\begin{equation}
\frac{d \langle h^2_0 \rangle}{dt}=-\frac{2\gamma}{t} \langle h^2_0
\rangle + 2F \langle h_0 \rangle + \frac{\epsilon t_0^\gamma}{L_0
t^\gamma},
\end{equation}
and the corresponding long time solution
\begin{equation}
\label{h02}
\langle h^2_0 \rangle=\frac{F^2}{(\gamma+1)^2}t^2 +
\frac{\epsilon t}{(\gamma + 1)L_0} \left( \frac{t_0}{t}
\right)^{\gamma},
\end{equation}
where the second summand in the right hand side of this equation
will be explicitly suppressed due to its subleading character, but
it will be taken into account implicitly later on in order to
construct Dirac delta functions out of infinite series. For the
other modes we find
\begin{equation}
\frac{d \langle h_n \rangle}{dt}=-\left( D\frac{n^2 \pi^2
t_0^{2\gamma}}{L_0^2 t^{2\gamma}}+\frac{\gamma}{t} \right) \langle
h_n \rangle,
\end{equation}
and we integrate it to find, in the long time limit, $\langle h_n
\rangle \to 0$. The correlation obeys the equation
\begin{equation}
\label{hmhn}
\frac{d}{dt} \langle h_m h_n \rangle=- \left[
D\frac{(m^2+n^2)\pi^2 t_0^{2\gamma}}{L_0^2
t^{2\gamma}}+\frac{2\gamma}{t} \right] \langle h_m h_n \rangle +
\frac{2 \epsilon t_0^\gamma}{L_0 t^\gamma}\delta_{mn},
\end{equation}
which solution is readily computable in terms of Misra functions
\cite{misra}; however its concrete form is complicated and not
particularly illuminating, so we will not reproduce it here. The
long time asymptotics of this solution depends on the value of
$\gamma$. For $\gamma < 1/2$ the interface dynamics is dominated by
diffusion, and one finds
\begin{equation}
\label{diffusion} \langle h_m h_n \rangle=\frac{2 \epsilon
\delta_{mn}L_0}{D(m^2+n^2)\pi^2} \left( \frac{t}{t_0}
\right)^\gamma.
\end{equation}
If $\gamma > 1/2$, then the interface dynamics is dominated by
dilution, and the asymptotic solution reads
\begin{equation}
\label{dilution}
\langle h_m h_n \rangle=\frac{2 \epsilon t_0
\delta_{mn}}{(1+\gamma)L_0} \left( \frac{t}{t_0} \right)^{1-\gamma}.
\end{equation}
Now we can reconstruct the first moments of the solution in
coordinate space. For the long time mean value we have
\begin{equation}
\langle h(x,t) \rangle= \frac{F}{\gamma + 1}t,
\end{equation}
and for the long time correlation when $x \neq x'$
\begin{equation}
\label{corr}
\langle h(x,t) h(x',t) \rangle=
\frac{F^2}{(\gamma+1)^2}t^2+ \frac{\epsilon L_0}{D \pi^2} \left(
\frac{t}{t_0} \right)^{\gamma} \sum_{n=1}^\infty \frac{1}{n^2} \cos
\left( \frac{n \pi x}{L_0} \right) \cos \left( \frac{n \pi x'}{L_0}
\right),
\end{equation}
if $\gamma < 1/2$, and
\begin{eqnarray}
\nonumber \langle h(x,t) h(x',t) \rangle &=&
\frac{F^2}{(\gamma+1)^2}t^2 + \frac{2 \epsilon t_0}{(\gamma + 1)L_0}
\left( \frac{t}{t_0} \right)^{1-\gamma} \sum_{n=1}^\infty \cos
\left( \frac{n \pi x}{L_0}
\right) \cos \left( \frac{n \pi x'}{L_0} \right) = \\
&=& \frac{F^2}{(\gamma+1)^2}t^2+\frac{\epsilon t_0}{\gamma +1}
\left( \frac{t}{t_0} \right)^{1-\gamma} \delta(x-x'),
\end{eqnarray}
if $\gamma > 1/2$ and asymptotically for long times, where we have
used the decomposition of the Dirac delta function in the same basis
as in Eq. (\ref{basis}) and we have implicitly taken into account
the subleading term in Eq. (\ref{h02}). Changing back to the
original Lagrangian coordinates, $y = x L(t)/L_0$, we find
\begin{equation}
\langle h(y,t) h(y',t) \rangle=\frac{F^2}{(\gamma+1)^2}t^2 +
\frac{\epsilon t}{\gamma +1} \delta(y-y'),
\end{equation}
the solution reduces to random deposition for long times. We thus
see that the surface stays uncorrelated if $\gamma
> 1/2$, while it becomes correlated when $\gamma < 1/2$. We can
further analyze the correlated phase summing Eq. (\ref{corr}) to
find
\begin{equation}
\langle h(x,t) h(x',t) \rangle= \frac{F^2}{(\gamma+1)^2}t^2 +
\frac{\epsilon}{12 D L_0} \left( \frac{t}{t_0} \right)^\gamma \left[
2L_0^2 -6L_0 \max(x,x') +3(x^2 +x'^2) \right],
\end{equation}
and in $y$ coordinates
\begin{equation}
\langle h(y,t) h(y',t) \rangle= \frac{F^2}{(\gamma+1)^2}t^2 +
\frac{\epsilon}{12 D L(t)} \left[ 2L(t)^2 -6L(t) \max(y,y') +3(y^2
+y'^2) \right].
\end{equation}
The one point correlation function reads
\begin{equation}
\langle h(x,t)^2 \rangle= \frac{F^2}{(\gamma+1)^2}t^2 +
\frac{\epsilon}{12 D L_0} \left( \frac{t}{t_0} \right)^\gamma \left[
2L_0^2 -6L_0 x +6 x^2 \right],
\end{equation}
or alternatively
\begin{equation}
\langle h(y,t)^2 \rangle= \frac{F^2}{(\gamma+1)^2}t^2 +
\frac{\epsilon}{6 D L(t)} \left[ L(t)^2 -3L(t) y +3 y^2 \right].
\end{equation}
Finally, we can calculate the height difference correlation
\begin{equation}
\left< \left[ h(y,t)-h(y',t) \right]^2 \right> = \frac{\epsilon}{2D}
\left| y-y' \right|,
\end{equation}
in agreement with the long time classical EW equation. Note that the
one point correlation function is not spatially homogeneous due to
the no-flux boundary conditions (periodic boundary conditions or
unbounded domains preserve the spatial homogeneity); to find a
coordinate independent result we define the surface width $W(t)$ as
in \cite{singha}
\begin{equation}
W(t)^2= \frac{1}{L_0} \int_0^{L_0} \left[ \langle h(x,t)^2
\rangle-\langle h(x,t) \rangle^2 \right] dx =
\frac{\epsilon}{12D}\left( \frac{t}{t_0} \right)^\gamma L_0=
\frac{\epsilon}{12D} L(t).
\end{equation}
From these formulas one finds that the width grows without achieving
saturation with an exponent $\beta_\infty = \gamma/2$, and its
square depends linearly on the substrate size, what allows to define
the roughness exponent $\alpha = 1/2$, as in the non-growing
substrate case, if we accept this definition despite the absence of
saturation. The marginal situation $\gamma = 1/2$ is characterized
by an equilibrium of diffusion and dilution, as can be seen by
regarding Eq. (\ref{hmhn}). In this case one can see that the
correlation length in the $x$ coordinates is $\lambda \approx
\sqrt{D t_0}$, what yields a correlation length in the $y$
coordinates $\Lambda \approx \sqrt{D t}$. This shows that the
fraction of interface that becomes correlated is
\begin{equation}
\frac{\Lambda(t)}{L(t)} \approx \frac{\sqrt{D t_0}}{L_0},
\end{equation}
revealing that the interface becomes globally correlated only if
diffusion is large enough, or alternatively if the initial system
size and growth rate are small enough; otherwise, the interface only
becomes partially correlated.

It is clear that Eq. (\ref{dilution}) is valid if the system is
observed from spatial distances $|x-x'| \gg (t/t_0)^{1/2-\gamma}$.
This might constitute a good approximation for the two point
correlation function, but it definitely breaks down when considering
the one point correlation. In this case we have to consider again
the solution of Eq. (\ref{hmhn}), but this time in the range $n,m
> (t/t_0)^{\gamma-1/2}$. In this scale the interface is again
dominated by diffusion and dilution may be disregarded, and we
obtain the same long time solution as in Eq. (\ref{diffusion}). The
asymptotic behavior is obtained summing this expression, but
including in the sum only the modes that contribute to the short
scale behavior $n,m > (t/t_0)^{\gamma-1/2}$. The result is
\begin{equation}
\langle h(x,t)^2 \rangle - \langle h(x,t) \rangle^2= \frac{\epsilon
L_0}{D \pi^2} \left( \frac{t}{t_0} \right)^\gamma
\sum_{n>(t/t_0)^{\gamma-1/2}} \frac{1}{n^2} \cos^2 \left( \frac{n
\pi x}{L_0} \right) \approx \frac{\epsilon L_0}{2 \pi^2 D} \left(
\frac{t}{t_0} \right)^{1/2},
\end{equation}
in the long time limit, where we have used the asymptotic expansion
of the Euler-Maclaurin formula representation of the series. This result
is valid as long as $0 \neq x \neq L_0$, otherwise
\begin{equation}
\label{boundary}
\langle h(0,t)^2 \rangle - \langle h(0,t)
\rangle^2=\langle h(L_0,t)^2 \rangle - \langle h(L_0,t) \rangle^2
\approx \frac{\epsilon L_0}{\pi^2 D} \left(\frac{t}{t_0}
\right)^{1/2}.
\end{equation}
Note that the behavior is now perfectly homogeneous, due to the
uncorrelated character of the interface in this case. The only
exception are the boundary points $x=0,L_0$, because they are
affected by growth only along one direction, and as a consequence
their auto-correlation is twice the value of the auto-correlation of
any other point not located at the boundary. These results hold
independently of the reference frame, be it Lagrangian or Eulerian,
precisely because the interface is uncorrelated. This allows us
calculating the height difference correlation function
\begin{equation}
\left< \left[ h(y,t)-h(y',t) \right]^2 \right> = \frac{\epsilon
L_0}{2 \pi^2 D} \left( \frac{t}{t_0} \right)^{1/2},
\end{equation}
when $y \neq y'$, $0 \neq y \neq L(t)$ and $0 \neq y' \neq L(t)$.
This shows the agreement with the short time classical EW equation.
Note that the dependence on the system size (both initial and time
dependent) is the same as in the static domain case, allowing the
definition of the roughness exponent, which takes on its classical
value.

In order to clarify the $\gamma>1/2$ dynamics more we will calculate
the two point correlation function in the long time limit and short
spatial scale $|x-x'| \ll (t/t_0)^{1/2-\gamma}$:
\begin{equation}
\langle h(x,t)h(x',t) \rangle -\langle h(x,t) \rangle^2=
|x-x'|\left( \frac{t}{t_0} \right)^\gamma \mathcal{F}_{EW}\left[
|x-x'|\left( \frac{t}{t_0} \right)^{\gamma-1/2} \right],
\end{equation}
or alternatively
\begin{equation}
\langle h(y,t)h(y',t) \rangle -\langle h(y,t) \rangle^2= |y-y'|
\mathcal{F}_{EW}\left[ |y-y'|\left( \frac{t}{t_0} \right)^{-1/2}
\right],
\end{equation}
where the scaling function reads
\begin{equation}
\mathcal{F}_{EW}(u)=\frac{\epsilon}{D} \left[ \frac{L_0}{2\pi^2 u}
\cos \left( \frac{u \pi}{L_0} \right)+\frac{1}{2\pi}
\mathrm{si}\left( \frac{u \pi}{L_0} \right) \right],
\end{equation}
where $\mathrm{si}(x)=-\int_x^\infty [\sin(s)/s] ds$ is the sine
integral. These results are again reminiscent of the scaling
behavior of the classical EW equation. It is remarkable how
translation invariance appeared in these formulas, despite the
presence of boundary conditions, as a consequence of decorrelation.

\section{General linear Langevin equation}
\label{general}

We now move to a more general situation in which we consider an
arbitrary diffusion operator of order $\zeta$ and an arbitrary
spatial dimension $d$, see Eq. (\ref{local}). From now on the
$d-$dimensional coordinates will be denoted ${\bf x} \to x$ and
${\bf y} \to y$ for simplicity. In this case, we can proceed
exactly in the same way as in the one-dimensional situation to
find, instead of Eq. (\ref{gamma}), the equation
\begin{equation}
\label{gammazeta}
\partial_t h=D \left( \frac{t_0}{t} \right)^{\zeta \gamma} |\nabla|^\zeta h
-\frac{d\gamma}{t}h+F+\left(\frac{t_0}{t}\right)^{d\gamma/2}\xi(x,t),
\end{equation}
where the fractional operator $|\nabla|^\zeta$ acts in Fourier space
as
\begin{equation}
\left( |\nabla|^\zeta h \right)_{\bf n} = -\frac{|{\bf n}|^\zeta
\pi^\zeta}{L_0^\zeta} h_{\bf n},
\end{equation}
where $(\cdot)_n$ denotes the corresponding Fourier transformed
quantity, and ${\bf n}=(n_1,\cdots,n_d)$. However, the solution of
a boundary value problem for an arbitrary fractional power of the
Laplacian is obtained not so straightforwardly. For some values of
the parameter $\zeta$ (such as for instance $\zeta \in (1,2)$)
this operator describes L\'{e}vy flights dispersal, for which even
the definition of boundary presents difficulties \cite{metzler}.
To overcome this pitfall we will decompose the solution using the
same basis as in the EW equation case, see Eq. (\ref{basis}), in
the case of a $d-$dimensional cubic box. This harmonic
decomposition implies the no flux boundary conditions not only for
$\zeta=2$, but for all positive even integer values of this
exponent. Indeed, for these values of $\zeta$ it is easy to check
that expanding the solution in the $d-$dimensional analog of
(\ref{basis}) is equivalent to prescribe the vanishing of all odd
order smaller than $\zeta$ derivatives of the solution at the
boundary. This fact allows us to propose a plausible definition of
solution to the no flux initial-boundary value problem
(\ref{gammazeta}) as the solution to this differential equation
which is expressed in terms of the $d-$dimensional version of
expression (\ref{basis}). This way the solution to the
initial-boundary value problem is defined in the context of
evolution semigroup theory as an initial value problem in a
prescribed functional domain \cite{evans}, which in this case is
generated by the selected eigenfunctions of the Laplacian, as in
Eq. (\ref{basis}).

In the present case the superuniversal threshold above which the
interface becomes uncorrelated turns out to be $\gamma=1/\zeta$.
For $\gamma$ greater than this value (and for large spatial scales
in case of the second moment), the two first moments of the
function height are given by
\begin{eqnarray}
\label{moment1}
\langle h(y,t) \rangle &=& \frac{F}{d\gamma +1}t, \\
\langle h(y,t) h(y',t) \rangle &=& \frac{F^2}{(d \gamma +1)^2}t^2
+ \frac{\epsilon t}{d \gamma + 1} \delta(y-y'). \label{moment2}
\end{eqnarray}
Lets now take a look of the one point correlation function. We can
proceed in the same way as in the previous section to find
\begin{equation}
\label{1pcorrz} \langle h(x,t)^2 \rangle - \langle h(x,t)
\rangle^2 \sim L_0^{\zeta-d} \left(\frac{t}{t_0}\right)^{(\zeta-d)
\gamma} \sum_{{\bf n}> (t/t_0)^{\gamma-1/\zeta}} \frac{1}{|{\bf
n}|^{\zeta}} \sim L_0^{\zeta-d}
\left(\frac{t}{t_0}\right)^{1-d/\zeta},
\end{equation}
by means of the asymptotic expansion of Euler-Maclaurin formula for
the series in the last step, and where
\begin{equation}
\sum_{{\bf n}} = \sum_{n_1} \cdots \sum_{n_d}.
\end{equation}
In Eq. (\ref{1pcorrz}) we have assumed $x=0$ for ease of analysis,
as we know that, in the limit of uncorrelated interface, all
points are equivalent up to a numerical prefactor present in the
boundary points \cite{fnote1}; the rough interface inequality
$\zeta
> d$ was assumed too. When this inequality is reversed $\zeta \le
d$, then the series does not converge. The situation is the same in
the case of a nongrowing domain, where for $\zeta \le d$ the
interface is flat or at most logarithmically rough, but this series
is divergent. Note that the one point correlation Eq.
(\ref{1pcorrz}) grows as $t^{2 \beta}$, where $\beta$ is the
corresponding growth exponent of the model without domain growth,
and as $L_0^{2\alpha}$ (and also as $L(t)^{2\alpha}$) for the
corresponding classical roughness exponent $\alpha$. These exponents
arise as a consequence of the local behavior of the model, which is
analogous to the static unbounded domain one, see below.

The behavior of the one point correlation function also helps
establishing some dynamical properties of the interface when $\gamma
< 1/\zeta$
\begin{equation}
\langle h(y,t)^2 \rangle - \langle h(y,t) \rangle^2 \sim
L_0^{\zeta-d} \left( \frac{t}{t_0} \right)^{(\zeta-d)\gamma}.
\end{equation}
As in previous cases, we note that there is no saturation, unlike in
the non-growing domain situation, for any $\gamma > 0$. One can
define the roughness exponent $\alpha$ from the second moment
dependence on the system size for long times. In the case of a
growing domain one would have in principle two possible choices: the
initial system size $L_0$ and the time dependent size $L(t)$. It
turns out that both yield the same value $\alpha= (\zeta-d)/2$, what
allows an unambiguous definition of this exponent. It is worthy
noting that this exponent is the exactly the same as in the regime
with $\gamma =0$, with the difference that this latter case is
characterized by the saturation of the fluctuations. If we allow the
definition of the long time growth exponent $\beta_\infty$ as the
power law dependence of the second moment on the temporal variable
for long times, we find $\beta_\infty=\gamma(\zeta-d)/2$. To
calculate an effective dynamic exponent $z_{\mathrm{eff}}$ we note
that the correlation length $\Lambda$ travels as $\Lambda(t) \approx
(Dt)^{1/\zeta}$, and so the correlated interface fraction at time
$t$ is
\begin{equation}
\frac{\Lambda(t)}{L(t)} \approx \frac{D^{1/\zeta}t_0^\gamma}{L_0}
t^{1/\zeta-\gamma},
\end{equation}
what leaves us
\begin{equation}
z_{\mathrm{eff}}=\frac{\zeta}{1-\zeta \gamma},
\end{equation}
if $\gamma < 1/\zeta$ and $\infty$ if $\gamma > 1/\zeta$. One sees
$z_{\mathrm{eff}}(\gamma=0)=\zeta$ and we recover the classical
case, and $\lim_{\gamma \to 1/\zeta}
z_{\mathrm{eff}}(\gamma)=\infty$, showing the limit in which the
interface becomes uncorrelated.

In the marginal situation characterized by $\gamma = 1/\zeta$
diffusion and dilution balance each other and so the resulting
dynamics is given by the concrete values of the equation parameters.
As we have seen, the effective dynamic exponent becomes divergent,
and the resulting fraction of correlated interface is
\begin{equation}
\frac{\Lambda(t)}{L(t)} \approx \frac{(Dt_0)^{1/\zeta}}{L_0},
\end{equation}
what shows that for large diffusion and small initial system size
and growth rate the interface becomes globally correlated.
Alternatively, for small diffusion and large initial system size and
growth rate the interface becomes only partially correlated.

As we have seen, the roughness exponent can be defined
$\alpha=(\zeta-d)/2$ independently of the value of $\gamma$, as in
the static domain case. For a supercritical $\gamma$, the long time
growth exponent reads $\beta_\infty = (1-d/\zeta)/2$, equalling the
growth exponent in the classical case. For any $\gamma \neq
1/(2\zeta)$ the inequality $\alpha \neq \beta_\infty
z_{\mathrm{eff}}$ holds, while the equality $\alpha = \beta_\infty
\zeta$ is true only if $\gamma \ge 1/\zeta$. On the other hand, we
have the alternative relation $\alpha=\beta_\infty z_\infty$, for
$z_\infty = \max \{1/\gamma,\zeta\}$. Indeed, there is a close
connection among $1/\gamma$ and the dynamic exponent, i. e., this
quantity describes quantitatively the speed at which correlations
propagate along the interface. This relation is further explored in
the following sections.

One can clarify things further by calculating the two point
correlation function for those bulk points that lie closer than the
correlation length $|x-x'| \ll (Dt)^{1/\zeta-\gamma}$. In $d=1$ we
obtain
\begin{equation}
\label{2pcorrzx} \langle h(x,t)h(x',t) \rangle -\langle h(x,t)
\rangle^2= \left[ |x-x'|\left( \frac{t}{t_0} \right)^\gamma
\right]^{\zeta-1} \mathcal{F}\left[ |x-x'|\left( \frac{t}{t_0}
\right)^{\gamma-1/\zeta} \right],
\end{equation}
or alternatively
\begin{equation}
\label{2pcorrz} \langle h(y,t)h(y',t) \rangle -\langle h(y,t)
\rangle^2= |y-y'|^{\zeta-1} \mathcal{F}\left[ |y-y'|\left(
\frac{t}{t_0} \right)^{-1/\zeta} \right],
\end{equation}
where the scaling function reads
\begin{equation}
\mathcal{F}(u)=\frac{\epsilon L_0^{\zeta-1}}{4 \pi^\zeta D}
u^{1-\zeta} \int_1^\infty s^{-\zeta} \cos \left(\frac{u \pi
s}{L_0}\right) ds,
\end{equation}
which integral can be considered as a trigonometric variant of the
Misra function \cite{misra}. These formulas have been found in the
long time limit after adiabatic elimination of highly oscillatory
functions (which results from a direct application of the
Riemann-Lebesgue lemma). Result (\ref{1pcorrz}) together with
(\ref{2pcorrz}) allows us to recover the classical critical
exponents $\alpha = (\zeta-1)/2$, $\beta=1/2-1/(2\zeta)$, and
$z=\zeta$, when we consider Lagrangian coordinates $y$ and
distances shorter than the correlation length. Note that Eqs.
(\ref{2pcorrzx}) and (\ref{2pcorrz}) only depend on $|x-x'|$ and
$|y-y'|$ respectively and thus they are translational invariant:
this is a consequence of decorrelation, what makes the interface
bulk behave similarly to the case of an unbounded domain for short
spatial scales. In this limit, boundary conditions are not
affecting the dynamics of bulk points, but boundary points show a
different prefactor as in the strictly local situation, see
(\ref{boundary}) and (\ref{1pcorrz}).

\section{Temporal correlations}
\label{temporal}

In order to calculate the temporal correlations we need to consider
EW dynamics Eq.(\ref{gamma}) in the short time limit, where the
growth exponent $\beta$ becomes apparent. The homogeneous solution
of its Fourier transformed representation Eq.(\ref{gammaf}) is
\begin{equation}
h_n (t)= \left(\frac{t}{t_0}\right)^{-\gamma} \exp \left[-\frac{n^2
\pi^2 D}{L_0^2}\frac{t_0^{2 \gamma} t^{1-2\gamma}-t_0}{1-2\gamma}
\right] h_n(t_0) \equiv G_n(t) h_n(t_0),
\end{equation}
that yields the following complete solution when the initial
condition vanishes:
\begin{equation}
h_n(t)=G_n(t) \int_{t_0}^{t} G_n^{-1}(\tau)
\left(\frac{t_0}{\tau}\right)^{\gamma/2} \xi_n(\tau) d\tau.
\end{equation}
The one point two times correlation function then reads
\begin{equation}
\label{corrf} \langle h_n(t)h_n(t') \rangle = \frac{2\epsilon}{L_0}
G_n(t) G_n(t') \int_{t_0}^{\min(t,t')} G_n^{-2}(\tau) \left(
\frac{t_0}{\tau} \right)^\gamma d\tau,
\end{equation}
and after inverting Fourier we arrive at the real space expression
\begin{equation}
\label{realseries}
\langle h(x,t)h(x,t') \rangle = \sum_{n=0}^\infty
\langle h_n(t) h_n(t') \rangle \cos^2 \left( \frac{n \pi x}{L_0}
\right).
\end{equation}
The propagator $G_n(t)$ suggests the scaling variable $v_n \sim
nt^{1/2-\gamma}$ in Fourier space, that corresponds to the real
space scaling variable $u \sim xt^{\gamma-1/2}$, as can be read
directly from the last equation. This again suggests the definition
of the effective dynamical exponent
$z_{\mathrm{eff}}=2/(1-2\gamma)$. If we express the correlation Eq.
(\ref{corrf}) for $t=t'$ in terms of the scaling variable $v_n$ (and
we refer to it as $C(v_n)$ multiplied by a suitable power of $t$)
and we introduce the "differential" $1 \equiv \Delta n \sim
t^{\gamma -1/2} \Delta v$, we can cast the last expression in the
integral form
\begin{equation}
\langle h(x,t)^2 \rangle -\langle h(x,t) \rangle^2 = t^{1/2}
\int_{v_1}^\infty C(v_n) \cos^2\left( \frac{v_n \pi u}{L_0} \right)
dv_n,
\end{equation}
where the series converges as a Riemann sum to the above integral
when
\begin{equation}
D t \ll (L_0^2+Dt_0){t^{2\gamma} \over t_0^{2\gamma}},
\end{equation}
or equivalently $t \ll t_c \sim L_0^{z_{\mathrm{eff}}}$, for $t_c$
being the time it takes the correlations reaching the substrate
boundaries, assuming that the substrate initial size is very
large. If $\gamma <1/2$, the whole substrate becomes correlated,
yielding a finite $t_c$; for $\gamma > 1/2$ the convergence of the
Riemann sum to the integral is assured for all times,
corresponding to the physical fact that the substrate never
becomes correlated. In front of the integral we find the factor
$t^{1/2}$, compatible with the growth exponent $\beta=1/4$, and
the integral can be shown to be absolutely convergent due to the
Gaussian dependence of $G_n(t)$ on $n$.

The general situation in which we deal with a $d-$dimensional
substrate and the diffusion is mediated by an operator of order
$\zeta$ can be constructed along the same steps. In this case the
propagator reads
\begin{equation}
\label{fouriermode} G_n(t)= \left(\frac{t}{t_0}\right)^{-d\gamma}
\exp \left[ -\frac{n^\zeta \pi^\zeta D}{L_0^\zeta} \frac{t_0^{\gamma
\zeta} t^{1-\gamma \zeta}-t_0}{1-\gamma \zeta} \right],
\end{equation}
suggesting that the scaling variables are $v_n \sim
nt^{1/\zeta-\gamma}$ and $u \sim xt^{\gamma - 1/\zeta}$, and a
definition for the effective dynamical exponent
$z_{\mathrm{eff}}=\zeta/(1-\gamma \zeta)$. We find again convergence
of the Riemann sum to an integral for any time if $\gamma >
1/\zeta$, and for short times
$$
D t \ll (L_0^\zeta+D t_0){t^{\zeta \gamma} \over t_0^{\zeta
\gamma}},
$$
if $\gamma < 1/\zeta$, in agreement with the expression for the
correlation time $t_c \sim L_0^{z_{\mathrm{eff}}}$. This integral is
again absolutely convergent as it decays superexponentially for
large values of the scaling variable $v_n$, leading to the result
\begin{equation}
\label{lrangle0}
\langle h(x,t)^2 \rangle -\langle h(x,t) \rangle^2
\sim t^{1-d/\zeta},
\end{equation}
in agreement with the classical growth exponent
$\beta=1/2-d/(2\zeta)$.

We are now in position to calculate the temporal auto-correlation
\begin{equation}
\label{temporalc} A(t,t') \equiv \frac{\langle
h(x,t)h(x,t')\rangle_0}{\langle h(x,t)^2 \rangle_0^{1/2} \langle
h(x,t')^2 \rangle_0^{1/2}} \sim
\left(\frac{\min\{t,t'\}}{\max\{t,t'\}}\right)^\lambda,
\end{equation}
where $\lambda$ is the auto-correlation exponent and $\langle
\cdot \rangle_0$ denotes the average with the zeroth mode
contribution suppressed, as in (\ref{lrangle0}). The remaining
ingredient is the correlation $\langle h(x,t)h(x,t')\rangle_0$.
Going back to Eq.(\ref{realseries}) we see that the Fourier space
scaling variable now reads
\begin{equation}
v_n=\left[\frac{t^{1-\gamma \zeta} +(t')^{1-\gamma \zeta} - 2
\tau^{1-\gamma \zeta}}{1-\gamma \zeta} \right]^{1/\zeta}n.
\end{equation}
If $\gamma < 1/\zeta$ the term $\max\{t,t'\}^{1-\gamma \zeta}$ is
dominant and the factor in front of the convergent Riemann sum reads
\begin{equation}
\max\{t,t'\}^{- d/\zeta} \min\{t,t'\},
\end{equation}
after the time integration has been performed and in the limit
$\max\{t,t'\} \gg \min\{t,t'\}$. In this same limit, but when
$\gamma > 1/\zeta$, the term $\min\{t,t'\}^{1-\gamma \zeta}$ becomes
dominant and the pre-factor reads
\begin{equation}
\max\{ t,t' \}^{-d\gamma} \min\{t,t'\}^{1- d/\zeta+d\gamma}.
\end{equation}
The resulting temporal correlation adopts the form indicated in the
right hand side of (\ref{temporalc}), where
\begin{equation}
\lambda = \left\{ \begin{array}{ll} \beta + d/\zeta &
\mbox{\qquad if \qquad $\gamma < 1/\zeta$}, \\
\beta +\gamma d & \mbox{\qquad if \qquad $\gamma
> 1/\zeta$}, \end{array} \right.
\end{equation}
or alternatively
\begin{equation}
\lambda= \beta + {d \over z_\lambda},
\end{equation}
where $\beta= 1/2-d/(2\zeta)$ and the new dynamical exponent is
defined as
\begin{equation}
z_\lambda = \min\{\zeta,1/\gamma\}.
\end{equation}
This last form is the natural generalization of the corresponding
one in \cite{kallabis}, and tells us that correlations are
propagated either by diffusion or dilution: the dominant mechanism
is chosen in each regime. We can extract more information about the
correlation function, as it decay properties
\begin{equation}
\langle h(x,t)h(x,t')\rangle_0 \sim \max\{t,t'\}^{-d/z}, \qquad
\mathrm{when} \qquad \max\{t,t'\} \to \infty,
\end{equation}
signaling that it decays to zero for long times. Also, the short
time behavior of the auto-correlation function is
\begin{equation}
A(t,t') \approx 1-R \left( 1-{\min\{t,t'\} \over \max\{t,t'\}}
\right)^{1-d/\zeta}, \qquad \mathrm{when} \qquad \max\{t,t'\}
\approx \min\{t,t'\},
\end{equation}
homogeneously in $\gamma$, where $R=R(\zeta,d,\gamma)$ is a
universal function of its arguments. This behavior is compatible to
the one found in the $\gamma=0$ case \cite{krug}. It indicates that
the short time properties of the auto-correlation are independent of
the substrate growth velocity, but the long time behavior is
influenced by the mechanism by which correlations are propagated, be
it diffusion or dilution.

\section{Persistence}
\label{persitence}

The persistence of a stochastic process denotes its tendency to
continue in its current state. When considering the dynamics of a
fluctuating interface, one refers to the persistence probability
$P_{+}(t_1,t_2)$ ($P_{-}(t_1,t_2)$) as the pointwise probability
that the interface remains above (below) its profile at $t_1$ up
to time $t_2>t_1$ \cite{kallabis,krug}. Herein, as in
\cite{singha}, we concentrate on the case in which the initial
profile is flat, and we suppress the contribution coming from the
zeroth mode as in the last section. For the stochastic growth
equations under consideration the symmetry $h_n \to -h_n$ for all
Fourier modes $n > 0$ holds, implying the equality $P_+ = P_-
\equiv P$. For long times $t_2 \gg t_1$ we have the power law
behavior \cite{kallabis,krug}
\begin{equation}
P(t_1,t_2) \sim (t_1/t_2)^\theta,
\end{equation}
defining the persistence exponent $\theta$. It was previously
calculated in the limit $\zeta \to \infty$ when $\gamma=0$
\cite{krug}
\begin{equation}
\theta \approx \frac{1}{2}+\frac{2\sqrt{2}-1}{2}\frac{d}{\zeta},
\end{equation}
up to higher order terms, and in this same limit when $d=1$ and
$\gamma=1$ \cite{singha}
\begin{equation}
\theta \approx \frac{1}{2}-\frac{1}{2\zeta},
\end{equation}
up to higher order terms. The goal of this section is to calculate
the persistence exponent $\theta$ in the limit $\zeta \to \infty$
for a finite, but otherwise arbitrary, value of $\gamma$. In order
to proceed with the calculation, we need to consider again the
normalized auto-correlation function, i. e., the left hand side of
(\ref{temporalc}). This time we will not focus on the limit $\max
\{t,t'\} \gg \min \{t,t' \}$, instead we will consider an arbitrary
relation among $t$ and $t'$. In this case we have \cite{foot}
\begin{eqnarray}
\nonumber \langle h(x,t)h(x,t')\rangle_0 \sim
\max\{t,t'\}^{-d\gamma} \min\{t,t'\} \left(\max\{t,t'\}^{1-\gamma
z}-\min\{t,t'\}^{1-\gamma z}\right)^{-d/z} \times
\\
\left[ 2\frac{ \min\{ t,t' \} \max\{ t,t' \}^{\gamma z} }{t
(t')^{\gamma z} + t^{\gamma z} t'} -1 \right]^{d/z}
{_2F_1}\left[\frac{\gamma d-1}{\gamma
z-1},\frac{d}{z};1+\frac{\gamma d-1}{\gamma z-1};2\frac{ \min\{ t,t'
\} \max\{ t,t' \}^{\gamma z} }{t (t')^{\gamma z} + t^{\gamma z}
t'}\right],
\end{eqnarray}
where ${_2F_1}(x_1,x_2;x_3;x_4)$ is Gauss hypergeometric function
\cite{gauss}. In order to derive the persistence exponent we
consider the auto-correlation function in logarithmic time
$T=\mathrm{ln}(t)$
\begin{equation}
A(T,T') \equiv \frac{\langle h \left(x,e^T \right)h \left(x,e^{T'}
\right)\rangle_0}{\langle h \left(x,e^T \right)^2 \rangle_0^{1/2}
\langle h \left(x,e^{T'} \right)^2 \rangle_0^{1/2}} \sim
e^{-(\beta+d\gamma)|T-T'|}, \qquad \mathrm{when} \qquad |T-T'| \to
\infty.
\end{equation}
Note that this is the correlation function for the normalized (to
unit variance) function height, which becomes stationary in the
logarithmic temporal variable. For short times we have
\begin{equation}
A(T,T') = 1 + \mathcal{O}\left(|T-T'|^{2 \beta}\right), \qquad
\mathrm{when} \qquad |T-T'| \to 0,
\end{equation}
homogeneously in $\gamma$. The first order term is a power $2
\beta=1-d/\zeta < 1$, classifying the process as a Slepian
non-smooth one \cite{slepian}. This fact together with the
asymptotics $A(T,T') \sim e^{-(1/2+d\gamma)|T-T'|}$ when $\zeta \to
\infty$ means that we can calculate the persistence exponent
$\theta$ perturbatively about $\theta=1/2+d\gamma$ for large $\zeta$
(and finite $d$ and $\gamma$) in the following fashion
\cite{singha,krug}
\begin{equation}
\theta \approx \left( {1 \over 2} +d\gamma \right) \left[
1-{1+2d\gamma \over \pi} \int_0^\infty \left\{
A(\tau)-e^{-(1/2+d\gamma)\tau} \right\} \left\{
1-e^{-(1+2d\gamma)\tau} \right\}^{-3/2} d\tau \right],
\end{equation}
for $\tau=|T-T'|$, yielding, in the limit $\zeta \to \infty$, the
result
\begin{equation}
\theta \approx \frac{1}{2} +d\gamma - \frac{d}{2\zeta},
\end{equation}
up to higher order terms. This last result is reminiscent of the
one obtained in \cite{singha}, but it is corrected by the effect
of dilution, as we shall discuss in the next section. We can see
that, in the limit considered, the interface is less persistent
than in the case of a static domain. This is so because in the
uncorrelated phase dilution acts as a relaxation mechanism on the
strictly local scale with a higher efficiency than diffusion. Note
that in the limit $\gamma \to 0$ one does not recover the static
domain result \cite{krug}. This is so because in this calculation
we have assumed $\gamma > 1/\zeta$, and so the vanishing $\gamma$
limit implicitly implies a faster vanishing $1/\zeta$ limit. As a
conclusion we find that when $\gamma \to 0$ the persistence
exponent $\theta \to 1/2$.

\section{Connection to Radial Growth}
\label{connection}

As mentioned in the Introduction, one of the characteristics of
radial growth is its growing domain interface. Herein, we will use
the results derived in previous sections to put former derivations
in radial geometry \cite{escudero,singha} in a broader context. In
this case as well dilution plays an important role on the
interface dynamics. As radial interfaces grow in time, the
interfacial matter becomes diluted among the new deposited matter
and the correlations transported simultaneously. Note that the
physical origin of dilution here is the same as in sec.
\ref{domains}, and thus it is not related to the surface
curvature.

The one-dimensional radial counterpart of the general linear
Langevin equation (\ref{gammazeta}) could be defined as
\cite{singha}
\begin{equation}
\frac{\partial r}{\partial t}= \gamma Ft^{\gamma-1} +
\frac{1}{r^\zeta} |\nabla_\theta|^\zeta r +
\frac{1}{\sqrt{r}}\eta(\theta,t),
\end{equation}
for the field $r(\theta,t)$, where reparametrization invariance
\cite{marsili} has been taken into account, but dilution has been
disregarded. Its analysis yields, for $\gamma > 1/\zeta$ and
performing a van Kampen system size expansion about the
homogeneously growing state, the long time large angular scale
correlation function \cite{escudero,escudero3}
\begin{equation}
\langle r(\theta,t)r(\theta',t) \rangle_0 \sim t^{1-\gamma} \,
\delta(\theta-\theta') \sim t \, \delta(s-s'),
\end{equation}
for $\gamma < 1$,
\begin{equation}
\langle r(\theta,t)r(\theta',t) \rangle_0 \sim \mathrm{ln}(t) \,
\delta(\theta-\theta') \sim t \, \mathrm{ln}(t) \, \delta(s-s'),
\end{equation}
for $\gamma=1$, and
\begin{equation}
\langle r(\theta,t)r(\theta',t) \rangle_0 \sim t_0^{1-\gamma} \,
\delta(\theta-\theta') \sim t^\gamma \, \delta(s-s'),
\end{equation}
for $\gamma > 1$, where $s-s' \sim t^\gamma (\theta-\theta')$ is
the arc-length scale. For $\gamma<1$ we recover the random
deposition correlation, while for $\gamma \ge 1$ we found, in the
arc-length variable, an average rapid roughening version of it,
this is:
\begin{equation}
\int_{s(\theta=0)}^{s(\theta=2\pi)} \langle r(s,t)r(s',t)
\rangle_0 \, ds \sim \left\{ \begin{array}{lll} t & \mbox{\qquad
if \qquad $\gamma <
1$}, \\ t \, \ln(t) & \mbox{\qquad if \qquad $\gamma = 1$}, \\
t^\gamma & \mbox{\qquad if \qquad $\gamma > 1$}.
\end{array} \right.
\end{equation}
This result emerges when the dilution term is not taken into
account. When we contemplate the effect of dilution, as in the
previous sections, we find a pure random deposition correlation in
Lagrangian coordinates
\begin{equation}
\langle r(s,t)r(s',t) \rangle_0 \sim t \, \delta(s-s'),
\end{equation}
homogeneously in $\gamma$ (provided $\gamma > 1/\zeta$), as in
(\ref{moment2}). Correspondingly, the prefactor of the Dirac delta
is $t^{1-\gamma}$ in the Eulerian setting. One can see that the
large scale results derived for radial interfaces
\cite{escudero,escudero3} are identical to the ones found here for
homogeneously growing domains, once dilution is introduced. These
results can be straightforwardly generalized to an arbitrary
dimension $d$. In this case one needs $d$ angles to parameterize the
interface in the Eulerian setting, that will lead to $d$ different
arc-lengths in Lagrangian coordinates and to a $d-$dimensional Dirac
delta specifying the spatial properties of the uncorrelated
interface. As a consequence one finds
\begin{equation}
\label{aroughness} \int_{S_t} \langle r({\bf s},t)r({\bf s}',t)
\rangle_0 \, d{\bf s} \sim \left\{ \begin{array}{lll} t &
\mbox{\qquad if \qquad $\gamma <
1/d$}, \\ t \, \ln(t) & \mbox{\qquad if \qquad $\gamma = 1/d$}, \\
t^{\gamma d} & \mbox{\qquad if \qquad $\gamma > 1/d$},
\end{array} \right.
\end{equation}
when dilution is not considered; ${\bf s}=(s_1,\cdots,s_d)$ is the
set of all arc-lengths and the integral extends to the whole domain.
If we contemplate dilution, then the resulting correlation in the
large spatial scale and for long times is pure $d-$dimensional
random deposition. The physical reason for this enhanced
stochasticity when dilution is not present is the following.
Dilution distributes the fluctuations along the interface at the
growth rate, keeping the total amount of noise constant. In its
absence, the interface is composed of unconnected sites, and new
ones are added in the process of domain growth. They act as
independent sources of noise, and so they contribute to augment the
overall fluctuations.

The study of radial growth performed in \cite{singha} on the
strictly local scale does not contemplate the effect of dilution
either. For the shake of completeness, and to facilitate
comparisons, we have derived the previous sections results
deliberately disregarding the effect of dilution. As expected, the
long time short spatial scale behavior is reminiscent of the
unbounded substrate situation, as happened in (\ref{2pcorrzx}) and
(\ref{2pcorrz}). On the contrary, the large spatial scale dynamics
shows a different phenomenology characterized by a particular
random deposition effective behavior. Equivalently, the long time
interval ($\max\{t,t'\} \gg \min\{t,t'\}$) asymptotics is affected
by substrate growth, and the absence of dilution modifies the
corresponding results. If $\gamma < 1/\zeta$, the correlation
function decays to zero at infinity as a power law
\begin{equation}
\langle h(x,t)h(x,t') \rangle_0 \sim (\max\{t,t'\})^{d \gamma -
d/\zeta}, \qquad \mathrm{when} \qquad \max\{t,t'\} \to \infty,
\end{equation}
while for $\gamma > 1/\zeta$ this correlation function approaches a
non-zero value. This fact is related to the behavior of the Fourier
modes (\ref{fouriermode}), which in the absence of dilution decay to
zero for $\gamma < 1/\zeta$ in the long time limit, while for
$\gamma > 1/\zeta$ they approach a non-zero value asymptotically in
time, analogously to the particular case analyzed in \cite{singha}.

The temporal auto-correlation function, in the absence of dilution,
is
\begin{equation}
\label{temporalc2} \frac{\langle h(x,t)h(x,t')\rangle_0}{\langle
h(x,t)^2 \rangle_0^{1/2} \langle h(x,t')^2 \rangle_0^{1/2}} \sim
\left(\frac{\min\{t,t'\}}{\max\{t,t'\}}\right)^\lambda,
\end{equation}
where
\begin{equation}
\lambda = \left\{ \begin{array}{ll} {1 \over 2} +\frac{d}{2\zeta} -d
\gamma & \mbox{\qquad if \qquad $\gamma < 1/\zeta$}, \\
{1 \over 2}-\frac{d}{2\zeta} & \mbox{\qquad if \qquad $\gamma
> 1/\zeta$}, \end{array} \right.
\end{equation}
or alternatively
\begin{equation}
\lambda= \beta + {d \over z_{\mathrm{eff}}},
\end{equation}
where $\beta= 1/2-d/(2\zeta)$ and
\begin{equation}
z_{\mathrm{eff}} = \left\{ \begin{array}{ll} \zeta/(1-\gamma
\zeta) & \mbox{\qquad if \qquad $\gamma < 1/\zeta$}, \\
\infty & \mbox{\qquad if \qquad $\gamma
> 1/\zeta$}. \end{array} \right.
\end{equation}
From these formulas one can clearly read that when dilution is
suppressed there is no mechanism for correlation propagation and
thus the system behaves as an effective particular random deposition
model in this limit. Similar information can be obtained from the
persistence exponent, obtained this time perturbatively about
$\theta=1/2$:
\begin{equation}
\theta \approx {1 \over 2} -\frac{d}{2 \zeta}, \qquad \zeta \to
\infty,
\end{equation}
generalizing the previous result \cite{singha}. This, together with
numerical simulations suggesting $\theta \approx 1/2$ almost
homogeneously in $\zeta$ \cite{singha}, reinforce the idea of an
effective particular random deposition behavior. However, for
$\zeta$ large enough $\theta < 1/2$, and in consequence the process
is more persistent than random deposition. This fact admits a
transparent physical explanation. For a static domain $\theta$
decreases for increasing $\beta$ \cite{kallabis}: the reason is that
in this case the exponent $\beta$ contains information on the
relaxation properties of the interface (through its dependence on
the dynamic exponent $z$). In this case relaxation is mediated by
diffusion, that connects the different interface points and thus
pushes the interface towards its mean value, diminishing the
persistence of the fluctuations. For a growing domain, dilution acts
as the relaxation mechanism when diffusion becomes inoperative (in
the uncorrelated phase). Suppressing dilution, there is no
relaxation mechanism left, following that the exponent $\beta$ only
contains information about the strictly local fluctuational
properties of the interface. Indeed, the interface variance grows as
$t^{2\beta}$ (see sec. \ref{general}), and so for smaller $\beta$ we
have weaker fluctuations intensity, implying a longer first passage
time. This implies in turn a smaller value for the persistence
exponent $\theta$, as persistence is nothing but a first passage
problem \cite{krug}. This explains how $\theta$ may increase for
increasing $\beta$, although understanding the whole numerical
sequence of values for the persistent exponent in \cite{singha}
would require a deeper analysis.

These last results for the temporal auto-correlation and persistence
show that in the uncorrelated phase dilution is the only responsible
for correlations propagation. Interestingly, the simulations
performed in \cite{singha} for the Eden model \cite{eden} show a
temporal autocorrelation function fully compatible with the one
described here for an uncorrelated interface in the absence of
dilution. The two persistence exponents, for above and below the
mean fluctuations (which are different for the less symmetric Eden
interface), are greater than $\theta = 1/2$, but are notably smaller
than the static domain ones \cite{singha}. This result is surprising
because for a radial cluster grown according to the Eden rules, to
which new cells are added at random positions on its interface,
dilution is expected to occur. It would be very interesting to
unveil the mechanism counterbalancing dilution in this case. It
might be a consequence of the particular way in which an Eden
cluster grows, or perhaps due to some possible nonlinear effect
acting on the interface as a consequence of geometry.

\section{Conclusions}
\label{conclusions}

In this work the dynamics of linear stochastic growth equations
whose domain size grows in time as power law $t^\gamma$ has been
studied. The growth index possesses one critical value
$\gamma=1/\zeta$, for $\zeta$ being the order of the diffusion
operator. If $\gamma < 1/\zeta$ the interface correlations are
propagated by means of diffusion at a faster speed than domain
growth, resulting in a fully correlated interface. The time it takes
correlations to travel the whole interface depends on the initial
substrate size $t_c \sim L_0^{z_\mathrm{eff}}$ for
$z_\mathrm{eff}=\zeta/(1-\gamma \zeta)$, or alternatively $t_c \sim
L(t_c)^\zeta$. The roughness exponent $\alpha$ can be defined from
the strictly local properties of the interface uniformly in the
growth index. This value of $\alpha$ is exactly the same one that is
obtained in the static domain case. For any $\gamma>0$ saturation
never occurs, and the interface width continues to grow for all
times with a long time growth exponent $\beta_\infty=\alpha \gamma$
when $\gamma < 1/\zeta$; note that in this regime $\beta_\infty <
\beta$ so there is partial saturation of the fluctuations. The
relation $\beta_\infty=\alpha \gamma$ also implies that correlations
travel like $t^\gamma$ in the long time regime, after they have
spread globally on the interface. Prior to that they propagate as
$t^{1/\zeta}$, because diffusion is a faster mechanism for
information transfer. Once they have reached the interface limits,
this transfer speed is limited by the slower process of domain
growth, resulting the stated exponents relation. This phenomenology
is independent of whether we contemplate dilution or not: it is a
strictly direct consequence of domain growth, not of dilution
(although dilution carries on information at the same velocity).

The regime in which $\gamma > 1/\zeta$ is characterized by a loss
of correlation along the interface. This translates into a delta
correlated spatial correlation for long times and large spatial
scales. The correlations for short spatial scales and time
intervals are reminiscent of the ones found in the case of a
static unbounded domain, revealing that diffusion is acting at
this level. On the other hand, large spatial scales and long time
intervals both reveal that the dominant mechanism for correlations
propagation is the dilution effect created by domain growth. By
means of dilution, correlations travel at the same speed at which
the domain grows, so a global correlation of the interface becomes
impossible. The situation is further clarified by the calculation
of the auto-correlation and persistence exponents. The
autocorrelation exponent $\lambda=\beta +d/\zeta$ for all $\gamma
< 1/\zeta$ (including $\gamma=0$) shows that one site interacts
with itself at former times by means of the growth process
(indicated by the first summand $\beta$) and with neighboring
sites by means diffusion (indicated by the second summand
$d/\zeta$), which is dominant in this regime. If $\gamma >
1/\zeta$, then the auto-correlation exponent reads $\lambda=\beta
+d\gamma$. This illustrates how, for fast domain growth, dilution
replaces diffusion and becomes responsible for the interaction
with the neighboring sites. A similar conclusion is reached by
analyzing the persistence of the surface fluctuations. While the
auto-correlation exponent yields information about the long time
interval asymptotics, the persistence exponent carries on
complimentary information obtained from averaging over all
possible time interval lengths. It reads $\theta=1/2 + d\gamma
-d/(2\zeta)$ when $\zeta \to \infty$ and $\gamma > 1/\zeta$, and
so it is greater than in the static domain situation, what implies
that the interface is less persistent. A reduced persistence is
associated with a stronger tendency to go back to the mean, which
is mediated by a stronger coupling with the neighboring points
through dilution, more efficient than diffusion in the limit
considered. Additionally, the persistence exponent increases with
the growth exponent: a higher $\beta$ corresponds to stronger
fluctuations and thus to a shorter first passage time. This is in
contrast to what happens in the $\gamma=0$ situation, where the
persistence exponent decreases for increasing $\beta$
\cite{kallabis}. The reason for this is that in this case $\beta$
contains information about the strength of the coupling among the
interface sites. Smaller $\beta$ implies a stronger coupling and a
correspondingly less persistent interface.

The crossover situation $\gamma=1/\zeta$ is characterized by a
nonuniversal behavior. There is a strong dependence on the
parameter values that enter in competition to yield the resulting
system dynamics. In the generic case $\gamma \neq 1/\zeta$, a
number of the results derived in this paper can be connected to
the Family-Vicsek ansatz \cite{family} by means of the simple
substitution $L \to L(t)$ taking into account the temporal
evolution of the system size directly in this ansatz. From the
results of this paper, one would expect this to be so in the
regime $\gamma < 1/\zeta$, or for $\gamma > 1/\zeta$ provided
short spatiotemporal distances are under consideration. When
$\gamma > 1/\zeta$, and for long time intervals and/or large
spatial scales one would perhaps expect a different result.
Indeed, in this regime correlations are propagated by means of
dilution, which becomes more effective than diffusion. However,
the surface dynamics is still well described by the Family-Vicsek
ansatz. To see this, consider for instance the unbounded domain
situation uniformly growing with a growth index $\gamma$. In this
case the Family-Vicsek ansatz tells us that the two points
correlation is:
$C(|x-x'|,t)=t^{2\beta}F(|x-x'|/t^{1/\zeta-\gamma})$ in Eulerian
coordinates $x$ and $2\beta=1-d/\zeta$ for the linear stochastic
growth equations considered here. For long times and $\gamma >
1/\zeta$ we find
$C(|x-x'|,t)=t^{1-d\gamma}t^{d\gamma-d/\zeta}F(|x-x'|/t^{1/\zeta
-\gamma}) \to t^{1-d\gamma}\delta(x-x')$ when $t \to \infty$, i.
e., we recover the standard random deposition correlation. This
result fully agrees with the correlations derived from the
stochastic growth equations which take into account dilution. On
the other hand, the correlations shown at the beginning of sec.
\ref{connection} were obtained suppressing the dilution term and
cannot be derived from the Family-Vicsek ansatz. So we see that
this ansatz implicitly takes into account dilution. This miracle
occurs because the Family-Vicsek ansatz neglects the memory with
respect to the initial condition at $t_0$. However, this memory
effect is present in the stochastic growth equations with no
dilution, leading to a different result \cite{escudero4}.
Introducing dilution asymptotically erases the memory with respect
to the initial condition, what implies in turn the coincidence of
the results from the stochastic growth equations and from the
Family-Vicsek ansatz. Of course, in this reasoning we have assumed
the rough interface inequality $\zeta > d$; otherwise the
appearance of non-universal anomalous dimensions is indeed
possible as we have shown for $\zeta = d$ in \cite{escudero3}.

One of the characteristics of radial growth is the growing domain
size. We have isolated this effect, what will hopefully allow a
better understanding of the dynamics of radial interfaces. In
order to facilitate the comparison with previous work we have
deliberately neglected the dilution term on the dynamics, although
this term arises naturally in the scenario considered. Our results
compared favorably and extended those of \cite{singha}. The
resulting analysis in the fast growth regime showed an interface
without relaxation mechanisms, and a corresponding temporal
auto-correlation function decaying to a non-zero value in the long
time limit. Once dilution is suppressed, and because diffusion is
inoperative in the large scale for $\gamma > 1/\zeta$, there is no
remaining coupling among the surface sites, as revealed by the
auto-correlation exponent $\lambda=\beta$ signalling that one site
only interacts with itself at different temporal points. The
information obtained from the persistence exponent is the same.
Note that the no dilution assumption allows growth characteristics
beyond what one would expect from the static domain results. For
instance, an average rapid roughening effect (not real rapid
roughening which would imply pointwise estimates out of reach in
this context) is possible for $\gamma \ge 1/d$, see Eq.
(\ref{aroughness}), because fluctuations are no longer being
distributed along the growing domain. Values of the persistence
exponent $\theta < 1/2$ strictly smaller than the random
deposition one are possible for large values of $\zeta$. Both
become impossible if the domain is non-growing or if we
contemplate dilution. Interestingly, simulations performed with
the Eden model (for which $\gamma = 1$) show that it behaves as if
dilution were not present \cite{singha}. Basically, this means
that an Eden cluster grows as a random deposition process, with
all its interface points being completely decorrelated (even at
the local scale for long time intervals), but with weaker surface
fluctuations as given by a smaller value of $\beta$ ($=1/3$ for
this particular model). The only dissenting factor is the
numerically measured value of the persistence exponents, greater
than the expected $\theta \approx 1/2$ \cite{singha}. Anyway,
these values are consistent with the rest of results, as they are
considerably smaller than the static domain ones \cite{kallabis},
implying a more persistent interface. They may be the consequence
of a KPZ nonlinearity (expected from the $\beta=1/3$ exponent)
acting on the interface, as it propagates correlations linearly in
time \cite{escudero}, and so it can compete with the linearly
growing Eden interface.

When studying a radial interface, it seems necessary to take into
account both reparametrization invariance \cite{marsili} and
dilution \cite{crampin}. The Eden model seems to be, in principle,
not an exception, as its interface grows by the addition of new
cells randomly placed on the cluster surface. However, numerical
results pointed to the fact that dilution is not operative in this
case \cite{singha}. It would be very interesting to understand what
mechanism is counterbalancing dilution in this model. Some
candidates are the presumed KPZ nonlinearity present in the Eden
surface dynamics, which is able to propagate correlations linearly
in time (exactly the same velocity at which dilution would operate
at the linearly growing Eden interface), or the nonlinearities
implied by reparametrization invariance.

\section*{Acknowledgments}

The author is grateful to Joachim Krug, Fabricio Maci\`{a} and
specially Juan J. L. Vel\'{a}zquez for helpful comments and
discussions. This work has been partially supported by the MICINN
(Spain) through Project No. MTM2008-03754.


\begin{thebibliography} {99}
\bibitem{fnote1} We conjecture that this factor is $2^b$, where $b$
is the number of faces of the hypercubic substrate the point in
question belongs to; for the origin, or any other corner, $b=d$.
\bibitem{foot} In this expression we have suppressed all constant
prefactors, including divergent terms of the form $1/(1-d\gamma)$.
Therefore the limit $\gamma \to 1/d$ must be taken with caution, and
one can not rely on this simplified formula. This includes as a
particular case the correlation calculated in \cite{singha}. The
rest of the results in this section do not depend on this
technicality.
\end{thebibliography}
\end{document}